\documentclass{cernrep} 
\usepackage{texnames}
\usepackage{caption}
\usepackage[bookmarks, colorlinks=true, linktoc=page, pdftex, linkcolor=black, citecolor=black, urlcolor=black]{hyperref}
\title{Multi-Particle Simulation Techniques}

\usepackage{fancyhdr}
\fancyhfoffset{4 mm}
\fancypagestyle{ARTTITLE}{%
\fancyhf{} 
\lhead{\small{Proceedings of the 2018 CERN--Accelerator--School course on\\
\it{Numerical Methods for Analysis, Design and Modelling of Particle Accelerators}, Thessaloniki, (Greece)}} 
\lfoot{Available online at \url{https://cas.web.cern.ch/previous-schools}}
\rfoot{\thepage\hspace*{3mm}}
 
}
\begin{document} 
\author {Ji Qiang}

\institute{Lawrence Berkeley National Laboratory, Berkeley, CA, USA}

\begin{abstract}
The nonlinear space-charge effects play an important role in
high intensity/high brightness accelerators.
These effects can be self-consistently studied using multi-particle simulations.
In this lecture, we will discuss the particle-in-cell
method and the symplectic tracking model for self-consistent multi-particle
simulations.
\end{abstract}


\maketitle 
\thispagestyle{ARTTITLE}
\section{Introduction}
The nonlinear space-charge effects from charged particle Coulomb interactions within
a bunched beam present a strong limit on beam intensity in
high intensity/high brightness accelerators by causing beam emittance
growth, halo formation, and even particle losses.
To model the details of beam distribution evolution, in the presence
of strong space-charge effects, one needs to solve the full Poisson-Vlasov
equations self-consistently.
The Poisson-Vlasov equations can be solved using a phase space grid-based method
or a multi-particle particle-in-cell (PIC) method. 
The grid-based method is effective in one or two dimensions~\cite{rob0}; but
for the three-dimensional system with six phase space variables, the grid-based method will
require an enormous amount of memory even for a coarse grid. Also, the grid-based method
may break down when very-small-scale structures form in the phase space. 
The particle-in-cell model is an efficient method to handle the space-charge effects 
self-consistently.
It has a much lower storage requirement and will not break down even when the phase space
structure falls below the grid resolution. 
It uses a computational grid to obtain the charge density distribution from
a finite number of macroparticles and solves the Poisson equation on the grid at each 
time step. The computational cost is linearly 
proportional to the number of macroparticles, which makes the simulation fast for
many applications. With the advance of computers, the PIC method has been implemented
on high performance parallel computers for large scale beam dynamics simulation.
The parallel PIC also provides a means of reducing fluctuations by enabling use of more particles 
and of improving spatial
resolution through increased grid density. It also dramatically reduces the computation time.
The PIC method has been widely used to study the dynamics
of high intensity/high brightness beams in accelerators~\cite{godfrey,friedman,jonesm,rob20,takeda,machida,jones,impact,track,toutatis,galambos,qin0,impact-t,amundson,opal,qiang09}.

\section{Particle-in-cell method for self-consistent multi-particle simulation}
In the particle-in-cell method, 
a number of macroparticles (i.e. simulation particles with the same
charge-to-mass ratio as the real charged particle) are generated from 
the Monte-Carlo sampling of a given initial distribution in phase space. 
The trajectories
of these charged macroparticles are tracked subject to both the external fields from 
the accelerating 
and focusing elements in the accelerator and from the self-consistent  
space-charge fields due to the Coulomb interactions among the charged particles.
\begin{figure}
   \centering
   \includegraphics*[angle=0,width=200pt]{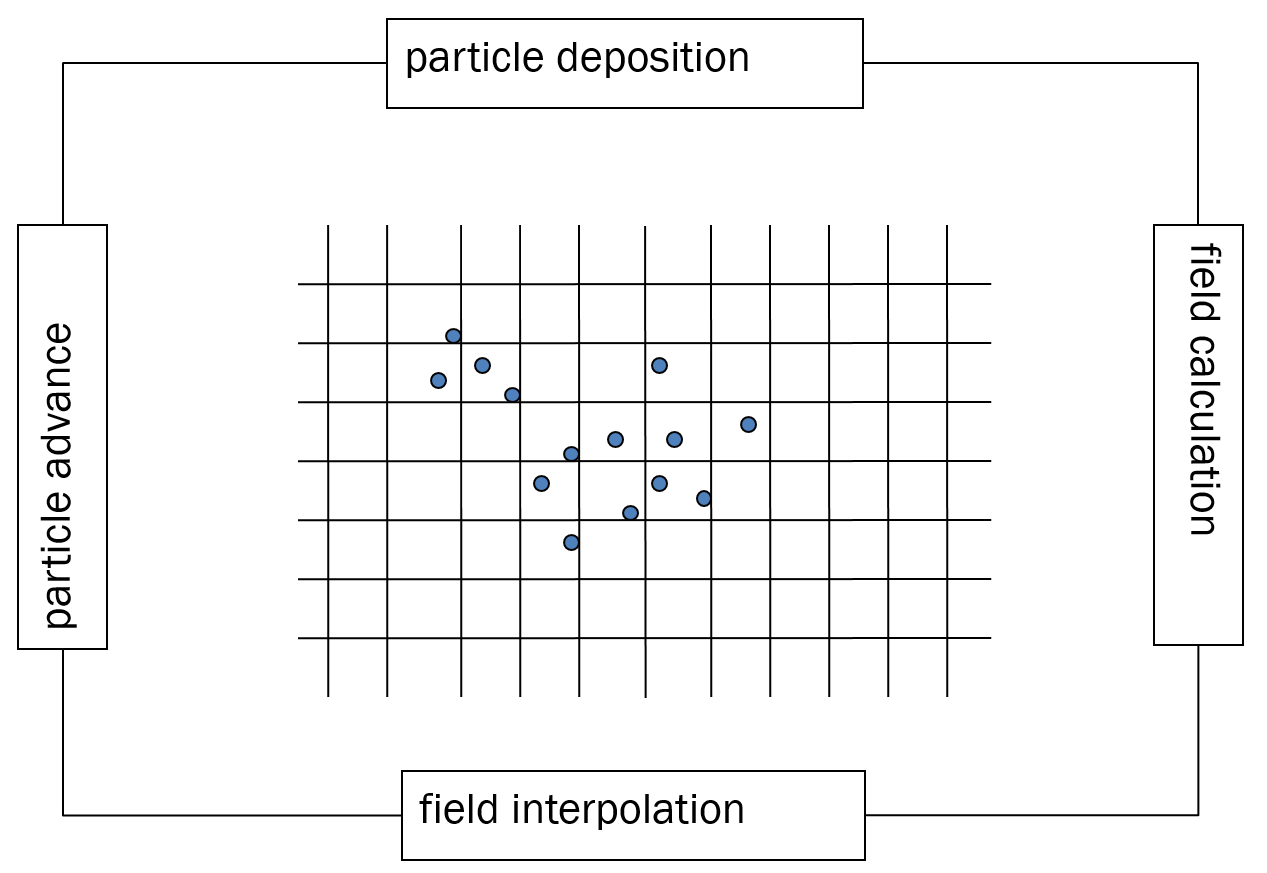}
   \caption{A schematic plot of a single step loop in the particle-in-cell 
	beam dynamics simulation.}
   \label{pic1step}
\end{figure}
Fig.~1 shows a schematic plot of a single step loop in the PIC beam dynamics simulation. 
It includes four stages: particle advance, charge deposition, self-consistent field calculation on mesh grid, and field interpolation. 
The particle deposition stage and the field interpolation stage are two symmetric stages that provide the information exchange between the point-like particle and the computational grid. 
Before the charge deposition, the particle coordinates are transformed into the moving beam frame from
the laboratory frame using the relativistic Lorentz transformation.
After the field interpolation, the fields are transformed back to the laboratory frame for particle momentum advance.
The field calculation stage solves the Poisson equation and calculates the space-charge fields
using the charge density distribution on the grid in the beam frame. 
The particle advance stage updates the particle position and momentum using the space-charge fields
and the external fields. In some numerical integrators such as the leap-frog integrator,
the update of the position and the update of the momentum can be done at separate sub-step locations within a step~\cite{leap}. 
This single step loop is repeated for a number of times in the accelerator beam
dynamics simulation.
In the following subsections, we will discuss these sub-steps in details. 

\subsection{Particle deposition and field interpolation}
\begin{figure}[!htb]
   \centering
   \includegraphics*[angle=0,width=200pt]{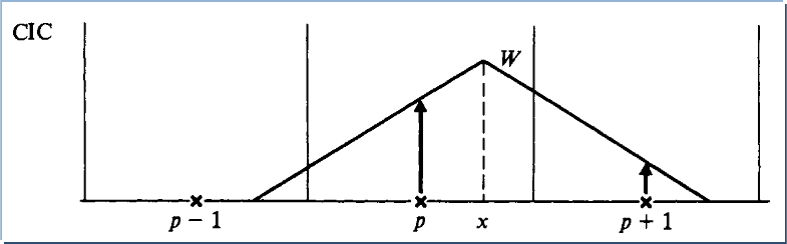}
   \includegraphics*[angle=0,width=200pt]{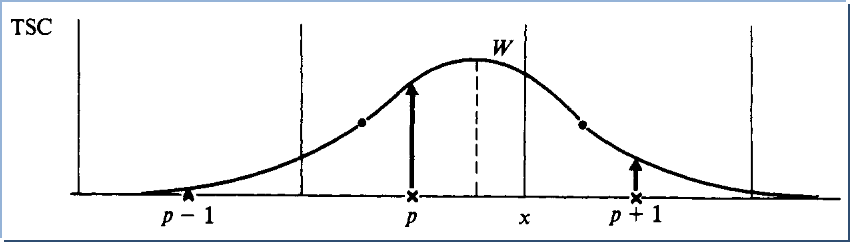}
	\caption{The CIC weight function (left) and the TSC weight function (right) used in
	charge density deposition~\cite{hockney}.} 
   \label{picdep}
\end{figure}

In the PIC method, the self-consistent space-charge fields are
calculated on a spatial grid using the charge density distribution on the grid at
every dynamically evolving step.
The density distribution on the grid is given by:
\begin{eqnarray}
	\rho_p(x_p) & = & \sum_i q_i w_{dep}(x_i - x_p) 
	\label{dep}
\end{eqnarray}
where $\rho_p$ is the charge
density on the spatial grid $p$, $q_i$ is the charge
	of the macroparticle $i$, $w_{dep}$ is the weight function for
	deposition, $x_i$ is the
spatial coordinates of particle $i$, and $x_p$ is the grid coordinate.
The solved space-charge fields on the grid are interpolated back to the
macroparticle position for momentum update. This is given by:
\begin{eqnarray}
	{\bf E}_i & = & \sum_p {\bf E}_p w_{int}(x_i - x_p) 
\end{eqnarray}
	where ${\bf E}_i$ is the electric
	field at the $i^{th}$ macroparticle location, ${\bf E}_p$
is the electric field at the grid point $p$,
and $w_{int}$ is the interpolation weight function from the grid point $p$ 
to the particle point $i$.
The use of the grid in the PIC method reduces the computational
cost of the space-charge field calculation compared with the direct 
point-to-point calculation that scales as the square of the
number of marcroparticles. 
The cost of field calculation
depends on the number of grid points, which is normally much less than
the number of macroparticles. 
The use of grid also provides smoothness to the shot noise and close
collision in the direct particle-to-particle calculation. 
From Eq.~\ref{dep}, by averaging the weighted charge of individual macroparticle, the charge
density distribution on the grid is smoother than the original particle
distribution. Meanwhile, the interactions among individual particles
become the interactions among spatial grid points.
In order to control the errors introduced in the deposition/interpolation
stage, one should choose the weight function so that 
the field error is small when the particle separations is large
compared with the mesh grid spacing.
Also, the charge assigned to the mesh from a particle and the field
interpolated to a particle from the mesh should vary smoothly as the
particle moves across the mesh. In many applications, in order to
avoid the self-force from a single particle on itself, the same weight
function for the deposition and the interpolation is used in
the PIC method, i.e. $w_{dep} = w_{int}$.
Two most widely used weight functions are the linear cloud-in-cell
(CIC) and the quadratic triangular shaped cloud (TSC) functions~\cite{hockney,birdsall}.
The weight function for the CIC is given as:
\begin{eqnarray}
w_p(x) & = & 1-|\frac{x-x_p}{h}| 
\end{eqnarray}
and
\begin{eqnarray}
	w_p(x) & = & \left \{ \begin{array}{ll} 
		\frac{3}{4} - (\frac{x - x_p}{h})^2, & |x - x_p| \leq h/2  \\
 \frac{1}{2}(\frac{3}{2}-\frac{|x-x_p|}{h})^2, & h/2 < |x - x_p| \leq 3/2 h \\
		 0 & {\rm otherwise}
                    \end{array}
                  \right.
\end{eqnarray}
for the TSC weight function. 
A schematic plot of those functions in one dimension are shown in Fig.~\ref{picdep}.
The linear CIC weight function involves two grid points in one dimension.
This weight function maintains the field continuity across the grid points.
The quadratic TSC function involves three grid points in one dimension.
The first derivative of the field value will be continuous using this weight function.
The higher order weight function can be constructed from the convolution
of the lower order weight function with the square nearest-grid-point
weighting function (also called top-hat function).
The higher order weight function is, the smoother the density function
will be, and the more computational cost it will take.

\subsection{Poisson solvers for space-charge field calculation}

Using the charge density on the grid, one can solve the Poisson equation 
subject to appropriate boundary conditions
in the beam frame to attain the self-consistent space-charge fields at each step.
In order to achieve reasonable simulation return time for practical applications
that involve thousands and even millions evolution steps,
the Poisson solver needs to be fast and computationally efficient.
Here, an efficient algorithm refers to the algorithm whose computational
cost scales linearly $(O(N))$ or log-linearly $(O(Nlog(N))$ with respect to
the number of unknowns to be solved.
Under different boundary conditions, different numerical algorithms
should be used to solve the Poisson equation effectively. 
In this lecture, we focus on two types of boundary conditions: one is the
open boundary condition, and the other is the finite domain boundary condition.

\subsubsection{FFT based Green's function method for the open boundary condition}
\begin{figure}[htb]
\centering
\includegraphics*[angle=0,width=80mm]{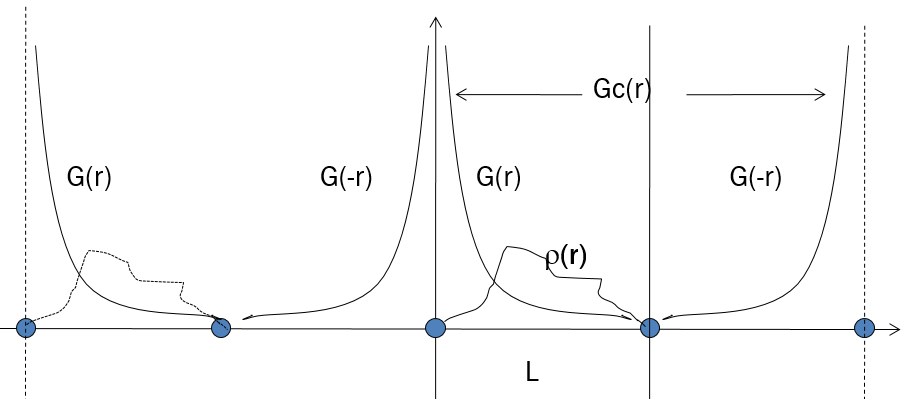}
\caption{A schematic plot of doubled computational domain
in one-dimensional case.}
\label{picgreen}
\end{figure}

The solution of the Poisson equation can be written as:
\begin{eqnarray}
\label{conv}
\phi(x,y,z) & = &
\frac{1}{4 \pi \epsilon_0}\int G(x,x',y,y',z,z') 
\rho(x',y', z') \ dx' dy' dz'
\end{eqnarray}
where $G$ is Green's function of the Poisson equation, $\rho$ is the 
charge density distribution function.
For a beam inside an accelerator, the
pipe aperture size is normally much larger than the transverse size of the
beam. In this case, an open boundary condition
can be assumed for
the solution of the Green's function in the above equation.
Here, the Green function
is given by:
\begin{eqnarray}
\label{green}
G(x,{x}',y,{y}',z,{z}') & = & \frac{1}{ \sqrt{(x-{x}')^2 + (y-{y}')^2
+(z-{z}')^2}}
\end{eqnarray}
Now consider a simulation of an open system where the 
computational domain containing the particles has a 
range of $(0,L_x)$, $(0,L_y)$ and $(0,L_z)$, and where each dimension is
discretized using $N_x$, $N_y$ and $N_z$ point,
from Eq.~\ref{conv}, the electric potentials on the grid can be
approximated as:
\begin{eqnarray}
\label{conv3d1}
{\phi}(x_i,y_j,z_k) & = & \frac{h_x h_y h_z}{4 \pi \epsilon_0}  
\sum_{i'=1}^{N_x} \sum_{j'=1}^{N_y} \sum_{k'=1}^{N_z} 
G(x_i-x_{i'},y_j-y_{j'},z_k-z_{k'}) 
{ \rho}(x_{i'},y_{j'},z_{k'})
\end{eqnarray}
where $x_i = (i-1)h_x$, $y_j = (j-1)h_y$, and $z_k = (k-1) h_z$.
The direct numerical summation of the above equation for all grid points 
can be very expensive and the computational cost scales as $N^2$, 
where $N=N_xN_yN_z$ is the total number of grid points.
Fortunately, this summation can be replaced by a summation
in a periodic doubled computational domain.
Fig.~\ref{picgreen} shows an illustrative plot of the doubled computational
domain in one-dimensional case.
In this periodic doubled computational domain, the original Green's function
in the negative domain, i.e. $G(-r)$, is mapped to the extended
domain following the periodic condition. The charge density in the extended
domain is set to zero. In this periodic system with a new periodic Green's
function and charge density, the summation can be done efficiently using
the FFT method whose computational cost scales as O(Nlog(N)). This new
summation yields exactly the same values 
as the original summation inside the original domain.

In the three-dimensional case, the cyclic convolution is given by~\cite{impact-t}:
\begin{eqnarray}
\label{conv3d2}
{ \phi}_c(x_i,y_j,z_k) & = & \frac{h_x h_y h_z}{4 \pi \epsilon_0} \sum_{i'=1}^{2N_x} \sum_{j'=1}^{2N_y} \sum_{k'=1}^{2N_z}
G_c(x_i-x_{i'},y_j-y_{j'},z_k-z_{k'}) 
{ \rho}_c(x_{i'},y_{j'},z_{k'})
\end{eqnarray}
where $i=1,\cdots,2N_x$, $j=1,\cdots,2N_y$, $k=1,\cdots,2N_z$ and
\begin{eqnarray}
{ \rho}_c(x_i,y_j,z_k) & = & \left\{ \begin{array}{r@{\quad:\quad}l}
{ \rho}(x_i,y_j,z_k) & 1\le i \le N_x; \ 1 \le j \le N_y; \ 1 \le k \le N_z \\
 0            & N_x < i \le 2 N_x \ or~ \ N_y < j \le 2 N_y \ or~ \ N_z < k \le 2 N_z
                              \end{array}
		      \right.  
\end{eqnarray}
{\small
\begin{eqnarray}
G_c(x_i,y_j,z_k) & = & \left\{ \begin{array}{r@{\quad:\quad}l}
G(x_i,y_j,z_k) & 1\le i \le N_x+1; \ 1 \le j \le N_y+1; \ 1 \le k \le N_z+1 \\
G(x_{2 N_x-i+2},y_j,z_k)   & N_x+1< i \le 2 N_x; \ 1 \le j \le N_y + 1; \ 1 \le k \le N_z+1 \\
G(x_i,y_{2 N_y-j+2},z_k)   & 1 \le i \le N_x+1; \ N_y+1 < j \le 2 N_y; \ 1 \le k \le N_z+1  \\
G(x_{2 N_x-i+2},y_{2 N_y-j+2},z_k)  & N_x+1 < i \le 2 N_x; \ N_y+1 < j \le 2 N_y; \ 1 \le k \le N_z+1  \\
G(x_i,y_j,z_{2 N_z -k +2} ) & 1\le i \le N_x+1; \ 1 \le j \le N_y+1;  N_z+1 < k \le 2 N_z \\
G(x_{2 N_x-i+2},y_j,z_{2 N_z -k +2})   & N_x+1< i \le 2 N_x; \ 1 \le j \le N_y + 1; N_z+1 < k \le 2 N_z \\
G(x_i,y_{2 N_y-j+2},z_{2 N_z -k +2})   & 1 \le i \le N_x+1; \ N_y+1 < j \le 2 N_y;  N_z+1 < k \le 2 N_z \\
G(x_{2 N_x-i+2},y_{2 N_y-j+2},z_{2 N_z -k +2})  & N_x+1 < i \le 2 N_x; \ N_y+1 < j \le 2 N_y;  N_z+1 < k \le 2 N_z \\
                              \end{array}
		      \right.  
\end{eqnarray}
}
\begin{eqnarray}
{ \rho}_c(x_i,y_j,z_k) & = & { \rho}_c(x_i+2(L_x+h_x),y_j+2(L_y+h_y),z_k+2(L_z+h_z)) \\
G_c(x_i,y_j,z_k) & = & G_c(x_i+2(L_x+h_x),y_j+2(L_y+h_y),z_k+2(L_z+h_z)) \ .
\end{eqnarray}
These equations make use of the symmetry of the Green function in Eq.~\ref{green}.
From the above definition, one can show that the cyclic convolution gives 
the same electric potential as the convolution Eq.~\ref{conv3d1} within the original 
domain, i.e.
\begin{eqnarray}
{ \phi}(x_i,y_j,z_k) & = & { \phi}_c(x_i,y_j,z_k) \hspace{1cm} for~ \ i=1, N_x; \ j=1,N_y; \ k=1,N_z \ .
\end{eqnarray}
The potential outside the original domain 
is incorrect but is irrelevant to the original physical domain. 
Since now both $G_c$ and ${ \rho}_c$ are periodic functions, the
convolution for ${ \phi}_c$ in Eq.~\ref{conv3d2} can be computed efficiently using the
FFT method.

In the above algorithm, both the Green function and the charge density 
distribution are discretized on the grid. For a beam with aspect ratio close to one, 
this algorithm works well. However, in some applications, for example,
during the emission of electrons out of the cathode,
the beam can have a very large transverse to longitudinal ratio. 
The typical transverse size is on the order of millimeters while the longitudinal size can be
about a few tens to hundred microns. Under this
situation, the direct use of the Green function on 
grid point is not efficient since it requires a large number
of grid points along the transverse direction in order to get
sufficient resolution for the Green function along that direction.
If we assume that the charge density function is uniform within
each cell centered at the grid point $(x_i,y_j,z_k)$, 
we can define an effective Green function as: 
{\small
\begin{eqnarray}
{\bar G}_(x_i-x_{i'},y_j-y_{j'},z_k-z_{k'}) & = & \int_{x_{i'}-h_x/2}^{x_{i'}+h_x/2} dx'
\int_{y_{j'}-h_y/2}^{y_{j'}+h_y/2} dy' \int_{z_{k'}-h_z/2}^{z_{k'}+h_z/2} dz' G(x_i-x',y_j-y',z_k-z')
\end{eqnarray}
}
This integral can be calculated analytically in a closed form~\cite{impacterr}:
\begin{eqnarray}
\int \int \int 
\frac{1}{\sqrt{x^2+y^2+z^2}} dxdydz & \doteq &
 -\frac{z^2}{2} \arctan(\frac{xy}{z\sqrt{x^2 + y^2 + z^2}}) - \frac{y^2}{2} \arctan(\frac{xz}{y\sqrt{x^2 + y^2 + z^2}})\nonumber \\
& & - \frac{x^2}{2} \arctan(\frac{yz}{x\sqrt{x^2 + y^2 + z^2}}) + yz\ln(x + \sqrt{x^2 + y^2 + z^2}) \nonumber \\
& & + xz\ln(y + \sqrt{x^2 + y^2 + z^2}) + xy\ln(z + \sqrt{x^2 + y^2 + z^2}) 
\end{eqnarray}

\begin{figure}[htb]
\centering
\includegraphics*[angle=0,width=75mm]{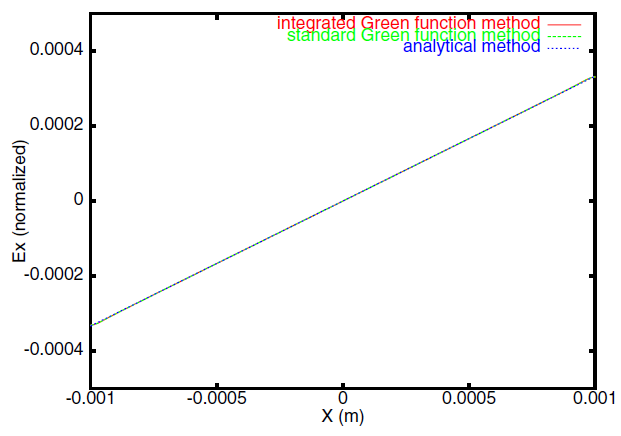}
\includegraphics*[angle=0,width=75mm]{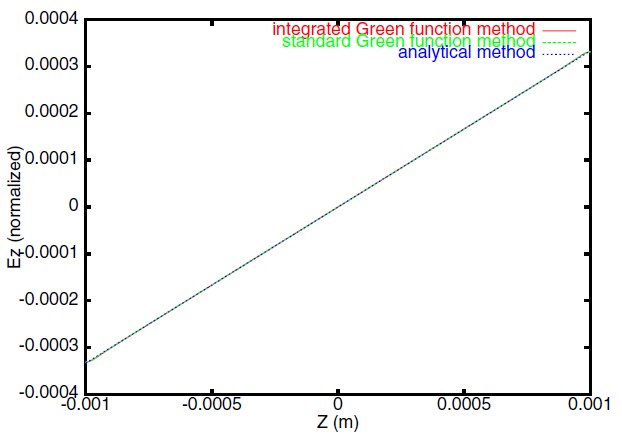}
\caption{Electric fields along x-axis (left) and along z-axis (right) of
a beam with aspect ratio one from solutions of the integrated 
Green function method, the standard Green function method, and the analytical
method.}
\label{picint1}
\end{figure}

\begin{figure}[htb]
\centering
\includegraphics*[angle=0,width=75mm]{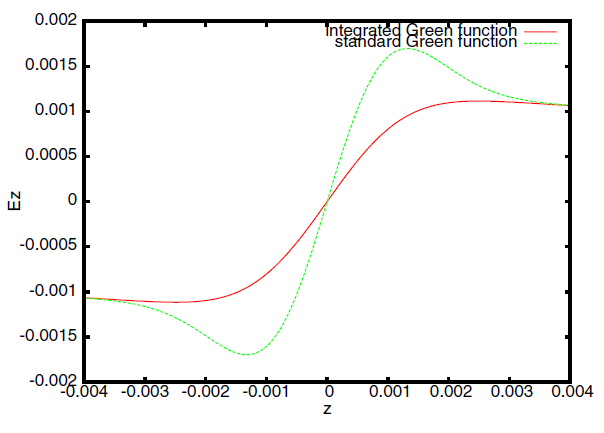}
\caption{Electric fields along z-axis of
a Gaussian beam with aspect ratio $30$ from solutions of the integrated 
Green function method and the standard Green function method.}
\label{picint}
\end{figure}

As a test of the above algorithm, we calculated the electric fields along the x-axis
and the z-axis of a charged beam with uniform density distribution in a spherical ball. 
The numerical results
from the integrated Green function together with
the solutions from the standard Green function method and the analytical 
solution are given in Fig.~\ref{picint1}.
With the aspect ratio one, 
all three solutions agree with each other very well.
For a Gaussian beam with an aspect ratio of $30$, the major discrepancy of the
electric field occurs around the core, which is given in Fig.~\ref{picint}.
These errors in the calculation of electric field for a large
aspect ratio beam using the standard
Green function method  
could significantly affect the accuracy of the beam dynamics simulation
inside the accelerator.

\subsubsection{Multigrid finite difference method for finite boundary condition}

In the application, when the effects from the boundary
wall become important, other efficient numerical methods should be
used to solve the Poisson equation.
For a simple regular boundary shape such as rectangular or round shape,
efficient Poisson solvers based on the spectral method have been developed~\cite{qiang01,qiang04,qiang16}. 
For a complex boundary geometry, the cut-cell multigrid finite difference method
can be used to handle the boundary condition~\cite{qiang06}.
In this method, the differential operator in the Poisson equation
is approximated by the difference operator
whose numerical accuracy depending on the separation of grid points. Away from the
boundary, the separation of grid points is normally uniform, i.e. uniform grid.
Near the boundary, nonuniform grid separations are used to fit the geometry of the
boundary. By replacing the differential operator with the difference operator,
the original Poisson equation is reduced to a group of linear algebraic equations.

For a one-dimensional Poisson equation:
\begin{eqnarray}
	\frac{\partial^2 \phi}{\partial x^2} & = & -\rho/\epsilon_0
\end{eqnarray}
Using a second-order finite difference approximation on a grid:
\begin{eqnarray}
	\frac{\partial^2 \phi}{\partial x^2} & \approx & \frac{1}{h^2}(\phi_{j+1}-2\phi_j
	+\phi_{j-1}) 
\end{eqnarray}
the Poisson equation is reduced to:
\begin{eqnarray}
	\frac{1}{h^2}(\phi_{j+1}-2\phi_j +\phi_{j-1}) & = & -\rho_j/\epsilon_0
\end{eqnarray}
where $j=1,\cdots,N$, and $N$ is total number of grid points.
The above linear algebraic equations can be rewritten in the matrix form:
\begin{eqnarray}
	\label{axb}
	A X & = & B
\end{eqnarray}
where $A$ is a sparse matrix, $X$ denotes the unknown electric potential $\phi$ 
on grid points,
and $B$ denotes the right hand side of the linear algebraic equations.
The direct solution of the above matrix equation using the Gaussian elimination
will take $O(N^3)$ operations, which is very inefficient.
For a sparse matrix $A$, the above equation is normally solved using
an iterative method whose computational cost scales as $O(mN)$. Here $m$
is the number of iterations. Using the iterative method, the above
linear algebraic equation can be rewritten as a recursive equation:
\begin{eqnarray}
\label{iteq}
	X^{i+1} & = & X^{i} + S(B-A  X^{i})
\end{eqnarray}
where $i=1,\ldots,m$ and $S$ is an approximation to $A^{-1}$ that can be computed quickly,
where $S=D^{-1}$
for the Jacobi method, $S=(D+L)^{-1}$ for the Gauss-Seidel
method, and $S=\omega(D+\omega L)^{-1}$ for the successive over relaxation (SOR) method,
in classical in classical iterative methods.
For a fast solver,
the number of iterations $m$ has to be reasonably small, 
i.e. the iteration has to converge to
the solution within a small number of iterations. However, for those classical
iterative methods, the convergence is slow.
These is because those classical iterative methods move information across
one grid at a time. It takes about $N^{1/2}$ steps to move information
across the grid. After a few iterations, the high frequency errors are
smoothed out while the low frequency errors decrease slowly.

\begin{figure}
\centering
	\includegraphics*[angle=0,width=80mm]{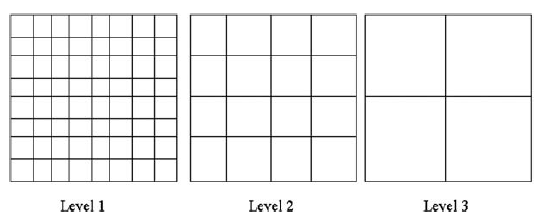}
\caption{Three-level discretization of a 2D computational domain.}
\label{multigrid}
\end{figure}

The multigrid method is an iterative method 
based on the concept of smoothing out numerical iteration errors on multiple
resolution scales. Instead of solving the original discrete Poisson
equation on one fixed mesh size, the multigrid method solves
the discrete Poisson problem on multiple levels of mesh size.
Fig.~\ref{multigrid} shows an example of three level discretization of a 2D
computational domain. 
As the discretization level increases, the
mesh size increases and the number of unknowns inside the
computational domain decreases exponentially.
At level three, only one unknown is left for
the Dirichlet boundary conditions, which can be solved directly.

\begin{figure}[htb]
\centering
\includegraphics*[angle=0,width=80mm]{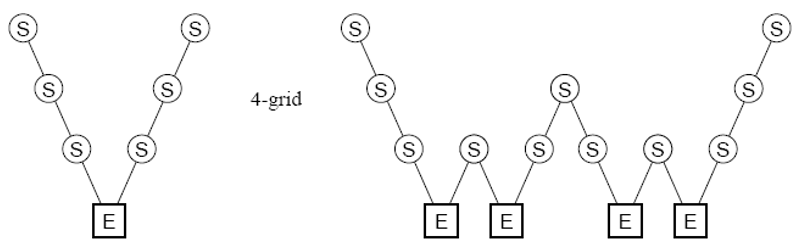}
\caption{A schematic plot of a V cycle and a W cycle iteration
structure in the multigrid algorithm using four grid levels~\cite{nr}.}
\label{vwcyc}
\end{figure}

The multigrid algorithm using two grid levels consists of five basic operations:
pre-smoothing, restriction, evaluation, prolongation, and post-smoothing.
During the pre-smoothing stage, an approximate solution of $X^{(i)}$,
$\tilde{X}^{(i)}$, is obtained using a
classical iterative method such as the Gauss-Seidel iteration:
\begin{eqnarray}
\tilde{X}^{(i)} & = & (D^{(i)}+L^{(i)})^{-1} U^{(i)} \tilde{X}^{(i)} + 
(D^{(i)}+L^{(i)})^{-1} B^{(i)}
\end{eqnarray}
where $A^{(i)}$ is the discrete Poisson operator at 
level $i$, $X^{(i)}$ denotes the unknown solution vector at the 
finest level or the unknown correction vector at other levels, and $B^{(i)}$ is
the source vector at the finest level or the residual vector at other
levels,
$D^{(i)}$, $L^{(i)}$ and $U^{(i)}$ correspond
to the main diagonal part, below diagonal part and above diagonal part
of the matrix $A^{(i)}$.
The residual vector $r^{(i)}$ at this level is calculated from the 
approximate solution as 
\begin{eqnarray}
r^{(i)} = B^{(i)} - A^{(i)}\tilde{X}^{(i)} 
\end{eqnarray}
These residuals are interpolated from the fine grid $i$ to the coarse grid $i+1$ using a restriction operator $\mathcal{R}$ 
to obtain the right hand side
of the Eq.~\ref{axb}:
\begin{eqnarray}
B^{(i+1)} = \mathcal{R} r^{(i)}
\end{eqnarray}
Here, a linear restriction operator $\mathcal{R}$ on a 2D grid is defined as:
\begin{eqnarray}
\mathcal{R} & = & \left( \begin{array}{ccc}
                  \frac{1}{16} & \frac{1}{8} & \frac{1}{16} \\
                  \frac{1}{8} & \frac{1}{4} & \frac{1}{8} \\
                  \frac{1}{16} & \frac{1}{8} & \frac{1}{16} 
                  \end{array} \right)
\end{eqnarray}
which corresponds to a bilinear nine-point interpolation scheme.
The evaluation operation on coarse grid $i+1$ will solve the
discrete Poisson equation for the correction vector 
through a direct or an iterative method. 
The obtained correction is reinterpolated back to the fine
grid $i$ from the coarse grid $i+1$ using a prolongation operator.
The improved solution on grid level $i$ is given by
\begin{eqnarray}
\tilde{X}^{(i)}_{new} = \tilde{X}^{(i)} + \mathcal{P} \tilde{X}^{(i+1)}
\end{eqnarray}
where $\mathcal{P}$ is the prolongation operator:
\begin{eqnarray}
\mathcal{P} & = & \left( \begin{array}{ccc}
                  \frac{1}{4} & \frac{1}{2} & \frac{1}{4} \\
                  \frac{1}{2} & 1 & \frac{1}{2} \\
                  \frac{1}{4} & \frac{1}{2} & \frac{1}{4} 
                  \end{array} \right)
\end{eqnarray}
which also corresponds to a bilinear interpolation scheme giving nonzero values
at nine grid points.
This new approximate solution 
is then used in the classical iterative method as a
post-smoothing stage to further improve the
accuracy of the solution. 
If the discrete equation on the coarse
grid $i+1$ can be solved using an evaluation operation, 
only two grid levels are used,
and the algorithm is referred to as two-grid method. 
If the solution on the coarse grid $i+1$ can not be easily attained,
the evaluation step can be replaced by more two-grid iterations. Depending on
how many two-grid iterations are used when each time 
the number of grid levels is increased by one, the multigrid
iteration can have a V cycle (one two-grid iteration is used) or
a W cycle (two two-grid iterations are used) structure~\cite{nr}.
Fig.~\ref{vwcyc} gives a schematic plot of a V cycle and a W cycle structure
with four grid levels.
Here, $S$ denotes a smoothing operation, $E$ denotes an evaluation operation,
each descending line $\backslash$ denotes a restriction operation, and each ascending
line $/$ denotes a prolongation operation. 

In the multigrid method, the iteration can start from the finest grid level
or start from the coarsest grid level. If a good initial 
guess of the solution is available, starting from the finest grid will be
an appropriate method. Otherwise, starting from the coarsest grid
will be more efficient since the solution on the coarsest grid
can be obtained from the direct evaluation and interpolated to the
next finer grid level. 
The interpolated solution on the finer grid level
is used to
start the smoothing operation at that level. After some V or W cycle
of iterations, the solution at that level
is further interpolated to next even finer grid level to start a new
smoothing operation and iteration cycle. Such a process is repeated
for a number of times until the finest grid level is reached. This type
of multigrid iteration is called full multigrid method or nested iteration. 

The multigrid method uses more grid levels than the conventional
single grid iteration methods. This seems to be more computationally
expensive than the single grid iteration.
However, by changing the
resolution of the discretization, i. e. scale of resolution, from one level
to the next level, 
the low frequency errors 
in the numerical residues of the iteration can be removed by a coarse grid iteration, 
while the high frequency errors can be removed on a fine grid iteration.
Therefore, the multigrid method uses much less number of iterations on
the finest grid level to obtain the converged solution 
than the single grid iteration method.
Most computational work is done on coarse grids with much less 
number of operations at each grid level compared
with the finest grid level iteration.
It has been shown that
the computational cost of this method scales linearly with the
number of grid points~\cite{wesseling}.
Hence, the multigrid iteration provides a much faster 
convergence than the classical
iterative methods such as the SOR method.

\subsection{Numerical integrators for particle advance}

With the self-consistent space-charge fields obtained from the solution of the Poisson
equation and the external fields, one can advance the particle using
a numerical integrator. 
This involves solving the
Lorentz force equations numerically. 
The Lorentz equations of motion for a charged particle subject to electric and
magnetic fields can be written as:
\begin{eqnarray}
	\frac{d {\bf r}}{d t} & = & \frac{{\bf p}c}{\gamma} \\
	\frac{d {\bf p}}{d t} & = & q(\frac{{\bf E}}{m c} + \frac{1}{m\gamma}
	        {\bf p}\times {\bf B})
\end{eqnarray}
where ${\bf r} = (x,y,z)$ denotes the particle spatial coordinates, ${\bf p}=
(p_{x}/mc,p_{y}/mc,p_{z}/mc)$ the particle normalized mechanic momentum,
$m$ the particle rest mass, $q$ the particle charge, 
$c$ the speed of light in vacuum,
$\gamma$ the relativistic factor defined by $\sqrt{1+{\bf p}\cdot {\bf p}}$,
$t$ the time, ${\bf E}(x,y,z,t)$ the electric field, and ${\bf B}(x,y,z,t)$ the magnetic field.

For a single step $\tau$, a second-order numerical integrator
for the above equation is given by:
\begin{eqnarray}
	\zeta (\tau) & = & {\mathcal M}(\tau) \zeta(0) \nonumber \\
    & = & {\mathcal M}_1(\tau/2) {\mathcal M}_2(\tau) {\mathcal M}_1(\tau/2) \zeta(0)
	+ O(\tau^3)
	\label{map}
\end{eqnarray}
The transfer map ${\mathcal M}_1(\tau/2)$ can be 
written as:
\begin{eqnarray}
	t(\tau/2) & = & t(0) + \frac{\tau}{2} \\
	{\bf r}(\tau/2) & = & {\bf r}(0) + \frac{\tau\bf p}{2\gamma}
\end{eqnarray}
The ${\mathcal M}_2(\tau)$ can have different second-order solutions depending
on different ways of approximation. In the Boris algorithm~\cite{boris}, 
${\mathcal M}_2(\tau)$ is given as:
\begin{eqnarray}
	\label{b1}
	{\bf p}_{-} & = & {\bf p}(0) + \frac{q {\bf E} \tau}{2mc} \\
	\gamma_{-} & = & \sqrt{1+{\bf p}_-\cdot {\bf p}_-} \\
	\label{b3}
	{\bf p}_+-{\bf p}_- & = & ({\bf p}_{+}+{\bf p}_{-}) \times 
\frac{q {\bf B}\tau}{2m\gamma_-}  \\
	\label{b4}
{\bf p}(\tau) & = & {\bf p}_+ + \frac{q {\bf E} \tau}{2mc} 
\end{eqnarray}
where ${\bf p}_+$ can be solved analytically from the linear equation
Eq.~\ref{b3}. 
The Boris algorithm is time-reversible and
has been widely used in numerical plasma simulations~\cite{birdsall}.
The particle momenta are updated using electric force in Eq.~\ref{b1} and
Eq.~\ref{b4}, and using magnetic force in Eq.~\ref{b3}.
The lack of direct cancellation between the electric fields and the magnetic fields
can introduce large error 
to simulate relativistic charged 
particles dynamics including space-charge effects, 
where the electric field and the magnetic field cancel each other
significantly in the laboratory frame and results in $1/\gamma^2$ decrease
of the transverse space-charge forces.

Another time-reversible solution for ${\mathcal M}_2(\tau)$ 
proposed in reference~\cite{vay} is given as:
\begin{eqnarray}
	\label{v1}
	\gamma_0 & = & \sqrt{1+{\bf p}\cdot {\bf p}} \\
	{\bf p}_{-} & = & {\bf p}(0) + \frac{q \tau}{2mc}({\bf E} 
	+ c{\bf p}/\gamma_0 \times {\bf B}) \\
	{\bf p}_{+} & = & {\bf p}_- + \frac{q {\bf E} \tau}{2mc} \\
	\gamma_1 & = & \sqrt{1+{\bf p}_+\cdot {\bf p}_+} \\
	{\bf t} & = & \frac{q{\bf B}\tau}{2m} \\
        \lambda & = & {\bf p}_{+} \cdot {\bf t} \\
	\sigma & = & \gamma_1^2-{\bf t}\cdot{\bf t} \\
        \gamma_2 & = & \sqrt{\frac{\sigma+
\sqrt{\sigma^2+4({\bf t}\cdot{\bf t} + \lambda^2)}}{2}}  \\
{\bf t}^* & = & {\bf t}/\gamma_2  \\
s & = & 1/(1+{\bf t}^* \cdot {\bf t}^*)   \\
	\label{v2}
{\bf p}(\tau) & = & s[{\bf p}_+ + ( {\bf p}_+ \cdot{\bf t}^*){\bf t}^* +
{\bf p}_+ \times {\bf t}^*]
\end{eqnarray}
This algorithm 
works well for charged particle tracking with large relativistic factor.
However, it is also mathematically more complicated than
the Boris algorithm and requires more numerical operations
than the Boris integrator. 

The source of error in the Boris
algorithm results from the lacking appropriate cancellation of the electric force
and the magnetic force.
This can be solved
by updating the momenta using both electric force and magnetic force
in the same step instead of separate steps. A
simple fast second-order integrator for the transfer
map ${\mathcal M}_2(\tau)$ that avoids the problem of the Boris algorithm was proposed
recently and is given as follows~\cite{qiang17b}:
\begin{eqnarray}
	\label{j1}
	{\bf p}_{-} & = & {\bf p}(0) + \frac{q}{mc} ({\bf E} + 
{\bf v}(0)\times {\bf B})\tau	\\
	\label{j2}
{\bf v}_+ & = & \frac{{\bf v}(0) + \bf{v}_-}{2} \\
	\label{j3}
{\bf p}(\tau) & = & {\bf p}(0) + \frac{q}{mc} ({\bf E} + 
{\bf v}_+\times {\bf B})\tau
\end{eqnarray}
where ${\bf v} = {\bf p}c/\gamma$.
This algorithm includes the direct cancellation of the electric force
and the magnetic force from the space-charge fields and works well
for large relativistic factor. It also has a simpler mathematical form
and
requires less numerical operations than the Boris integrator
and the integrator in Eqs.~\ref{v1}-\ref{v2}. This algorithm converges to the solution of the above integrator
if one repeats Eqs.~\ref{j1}-\ref{j3} many times. However, this is not 
necessary since 
the iteration does not increase the order of accuracy of the algorithm. 
It is shown in the following examples
that the new numerical integrator agrees with the integrator
in Eqs.~\ref{v1}-\ref{v2} very well.

\subsubsection{Benchmark of the numerical integrators}

The above second-order numerical integrator 
were benchmarked
using the following numerical example.
In this example, we considered an electron moving inside
the static electric and magnetic fields generated by a
co-moving positron beam. These fields are
given as:
\begin{eqnarray}
	E_x  & = & E_0 x \gamma_0 \\
	E_y  & = & E_0 y \gamma_0 \\
	E_z  & = & 0 \\
	B_x  & = & -E_0 y\gamma_0 \beta_0/c \\
	B_y  & = & E_0 x \gamma_0 \beta_0/c \\
	B_z  & = & 0 
\end{eqnarray}
where $\gamma_0$ is the relativistic factor of the moving positron beam,
$\beta_0 = \sqrt{1-(1/\gamma_0)^2}$, and the constant $E_0 = 9 \times 10^6 V/m^2$.
The above fields correspond to 
the space-space fields generated by the co-moving
infinitely long transversely uniform cylindrical positron beam.

\begin{figure}[!htb]
   \centering
   \includegraphics*[angle=0,width=200pt]{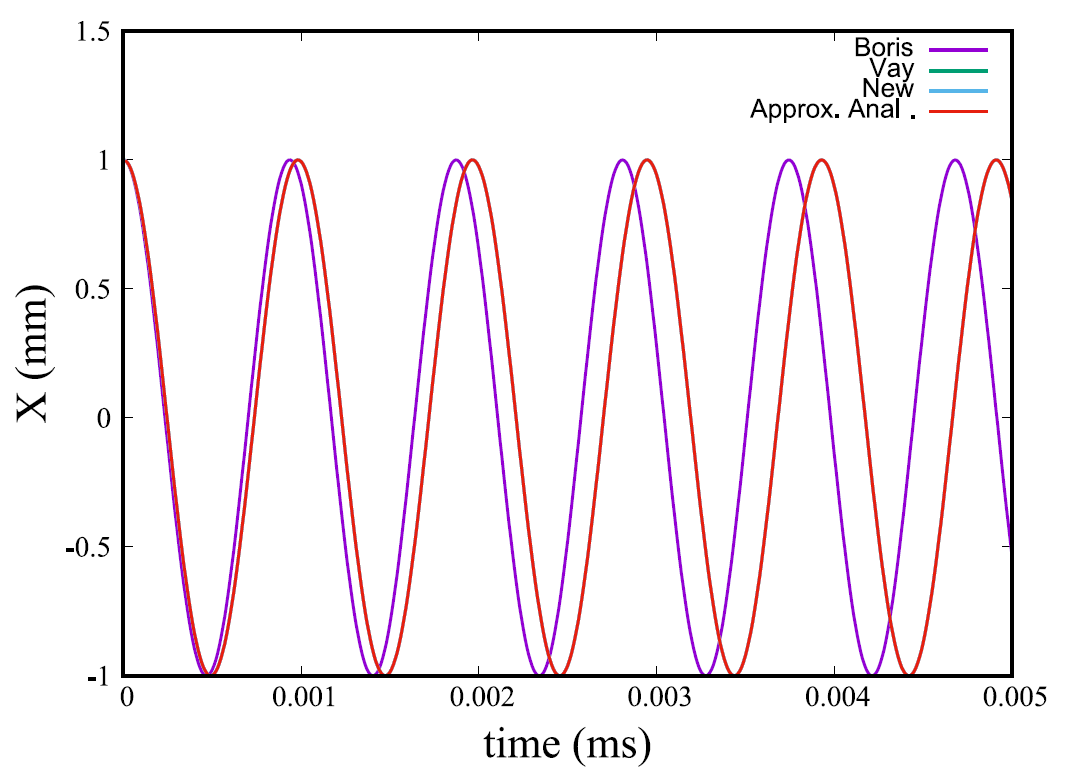}
   \caption{Particle $x$ coordinate evolution as a function of time from the
   Boris integrator (magenta), the Vay integrator (green),
   the new integrator (blue), and the analytical solution (red) for an electron
   with $100$ MeV kinetic energy.}
   \label{fig3}
\end{figure}

\begin{figure}[!htb]
   \centering
   \includegraphics*[angle=0,width=200pt]{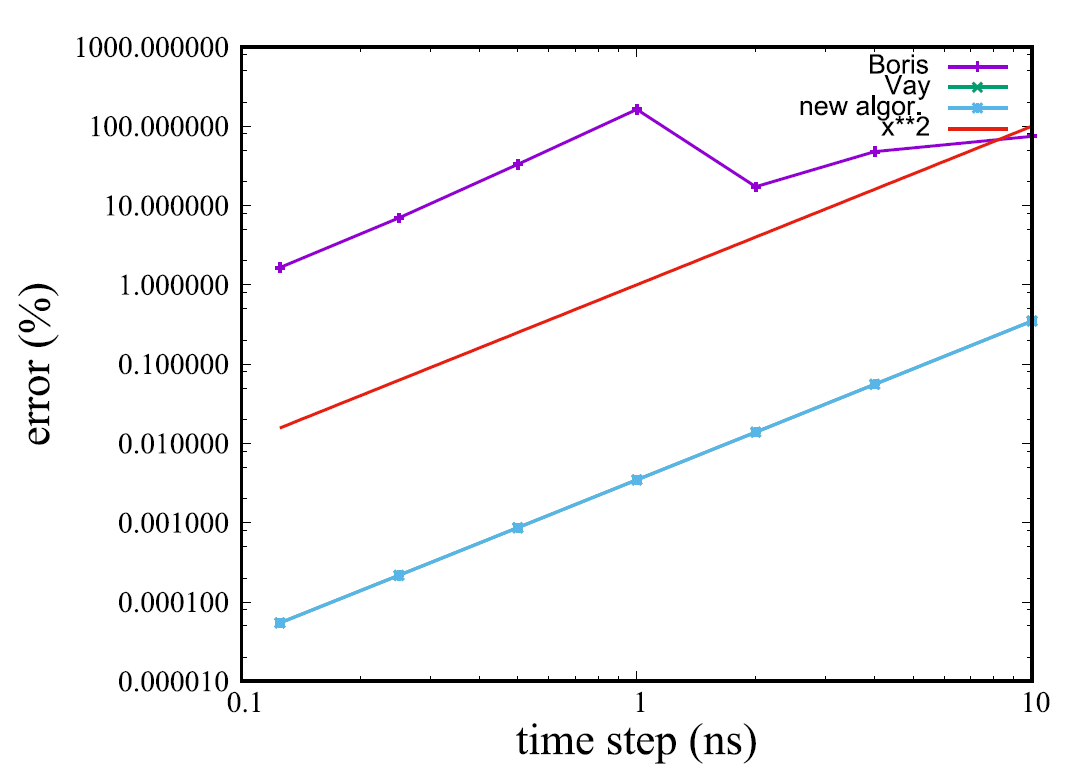}
      \caption{Relative numerical errors at the end of the above integration
	   as a function of step size from
   the Boris integrator (magenta), the Vay integrator (green),
   and the new integrator (blue) for an electron
   with $100$ MeV kinetic energy. A quadratic monomial function is also plotted here (red).}
   \label{fig4}
\end{figure}

We assumed that both the initial electron and the co-moving positron beam
have a kinetic energy of $100$ MeV.
Fig.~\ref{fig3} shows the electron $x$ coordinate evolution as a function of time from the
   Boris integrator (magenta), the integrator proposed by Vay (green),
   the new integrator (blue) with a step size of $1$ ns (around $0.001$ oscillation period), and the analytical solution (red).
Here, the analytical solution is obtained in the
co-moving frame without including the relativistic effects and then
Lorentz transformed back to the laboratory frame.
The analytical solution for the $x$ trajectory starting with initial 
$0$ momentum is given as:
\begin{eqnarray}
	x(t) & = & x_0 \cos(\sqrt{q E_0/m}/\gamma_0 t)
\end{eqnarray}
where $x_0 = 1$ mm is the initial horizontal position.
It is seen that after one oscillation period, the solution from
the Boris integrator starts to deviate from the other solutions while
the other three solutions are still on top of each other.
Fig.~\ref{fig4} shows the relative numerical errors at the end
of the above integration as a function of time step size from
the Boris integrator, the Vay integrator, and the new integrator.
As expected, all three second-order accuracy numerical integrators converge
quadratically with respect to the time step size.
However, the errors from the Boris integrator are about $4$ orders of magnitude
larger than those from the other two integrators. 

\begin{figure}[!htb]
   \centering
   \includegraphics*[angle=0,width=200pt]{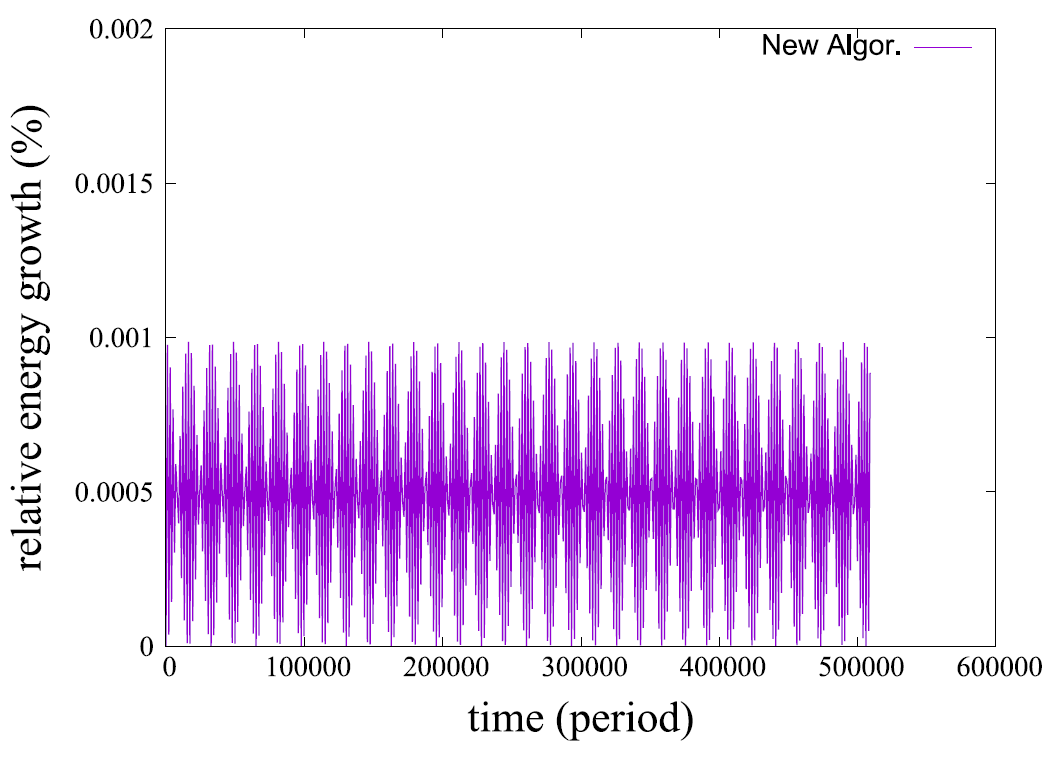}
   \caption{Relative kinetic energy growth evolution of an 
   initial $100$ MeV electron.}
   \label{fig5}
\end{figure}

\begin{figure}[!htb]
   \centering
   \includegraphics*[angle=0,width=200pt]{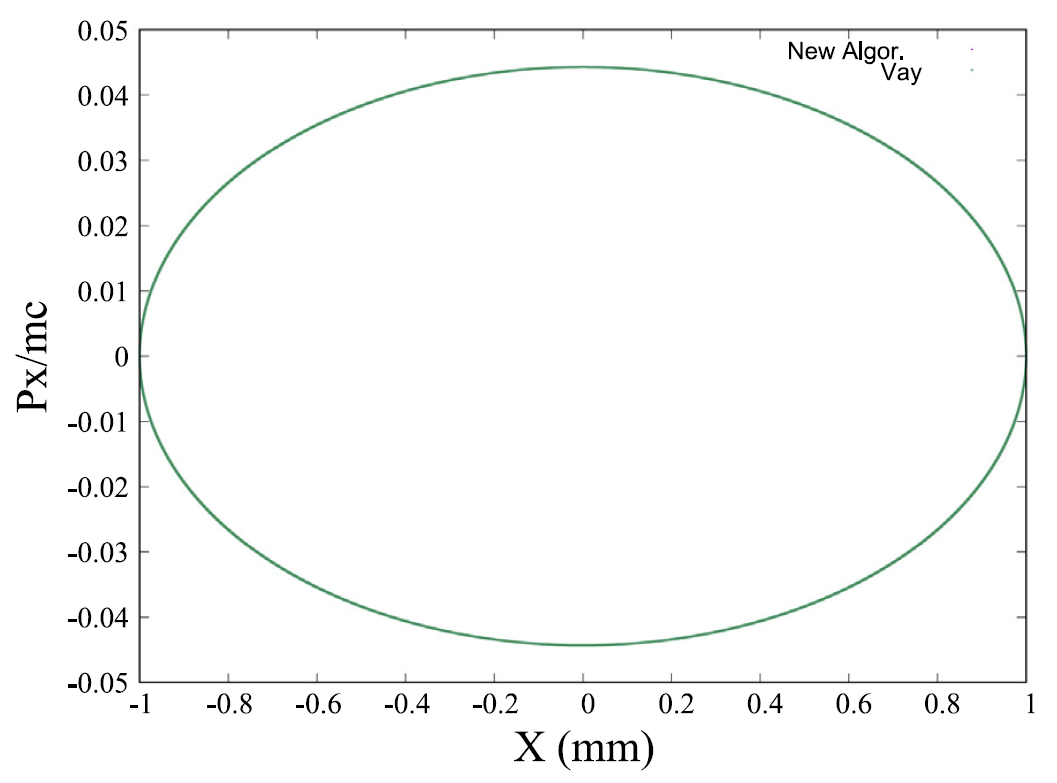}
   \caption{Electron phase space trajectory from
   the new integrator and the Vay integrator.}
   \label{fig6}
\end{figure}

In the second example, we tracked a $100$ MeV
electron in the above electric and magnetic fields
for more $500,000$ periods using the new
second-order numerical integrator with time step size of $100$ ns. 
The relative
kinetic energy growth as a function of time is shown in
Fig.~\ref{fig5}. It is seen that except the
oscillation from energy exchange, there is no steady state
secular energy increase or decrease resulting from numerical
heating or damping of the proposed new integrator.
Fig.~\ref{fig6} shows the phase space trajectory
of the electron from the proposed new algorithm and from the Vay
algorithm. It is seen that both integrators agree with each other very
well. The phase space structure is well preserved
after $500,000$ periods. 

\section{Symplectic self-consistent space-charge tracking models}

The above grid based, momentum conserved, particle-in-cell method does not 
satisfy the symplectic condition of classic multi-particle dynamics. 
Violating the symplectic condition
in multi-particle tracking might not be an issue in a single pass system such as
a linear accelerator. 
But in a circular accelerator, 
violating the symplectic condition
may result in undesired numerical errors in the long-term tracking simulation.
Recently, symplectic multi-particle model was proposed for
self-consistent space-charge simulations~\cite{qiang17,qiang18}.

In the accelerator beam dynamics simulation, for a multi-particle system with $N_p$ charged particles subject to both a space-charge self field
and an external field, an approximate Hamiltonian of the system can 
be written as~\cite{rob1}:
\begin{eqnarray}
	H & = & \sum_{i=1}^{N_p} {\bf p}_i^2/2 + \frac{1}{2} \sum_{i=1}^{N_p} 
	\sum_{j=1}^{N_p} q \bar{\varphi}({\bf r}_i,{\bf r}_j)
	+ \sum_{i=1}^{N_p} q \psi({\bf r}_i)
	\label{htot}
\end{eqnarray}
where $H({\bf r}_1,{\bf r}_2,\cdots,{\bf r}_{N_p},{\bf p}_1,{\bf p}_2,\cdots,{\bf p}_{N_p};s)$ denotes the Hamiltonian of the system using distance $s$ as an independent variable, 
$\bar{\varphi}$ is the space-charge interaction potential (including both the direct
electric potential and the longitudinal vector potential) 
between the charged
particles $i$ and $j$ (subject to appropriate boundary conditions), $\psi$ denotes the
potential associated with the external field,
${\bf r}_i=(x_i,y_i,\theta_i=\omega \Delta t)$ denotes the normalized canonical
spatial coordinates of particle $i$, ${\bf p}_i=(p_{xi},p_{yi},p_{ti}=-\Delta E/mc^2)$ the normalized canonical momentum
coordinates of particle $i$, and $\omega$ the reference angular frequency,
$\Delta t$ the time of flight to location $s$, $\Delta E$ the energy
deviation with respect to the reference particle, $m$ the rest mass of the
particle, and $c$ the speed of light in vacuum.
The equations governing the motion of individual particle
$i$ follow the Hamilton's equations as:
\begin{eqnarray}
	\frac{d {\bf r}_i}{d s} & = & \frac{\partial H}{\partial {\bf p}_i} \\
	\frac{d {\bf p}_i}{d s} & = & -\frac{\partial H}{\partial {\bf r}_i} 
\end{eqnarray}
Let $\zeta$ denote a 6N-vector of coordinates,
the above Hamilton's equation can be rewritten as:
\begin{eqnarray}
	\frac{d \zeta}{d s} & = & -[H, \zeta] 
\end{eqnarray}
where [\ ,\ ] is the Poisson bracket. A formal solution for the above equation
after a single step $\tau$ can be written as:
\begin{eqnarray}
	\zeta (\tau) & = & \exp(-\tau(:H:)) \zeta(0)
\end{eqnarray}
Here, we have defined a differential operator $:H:$ as $: H : g = [H, \ g]$, 
for arbitrary function $g$. 
For a Hamiltonian that can be written as a sum of two terms $H =  H_1 + H_2$, an approximate
solution to above formal solution can be written as~\cite{forest1}
\begin{eqnarray}
	\zeta (\tau) & = & \exp(-\tau(:H_1:+:H_2:)) \zeta(0) \nonumber \\
  & = & \exp(-\frac{1}{2}\tau :H_1:)\exp(-\tau:H_2:) \exp(-\frac{1}{2}\tau:H_1:) \zeta(0) + O(\tau^3)
\end{eqnarray}
Let $\exp(-\frac{1}{2}\tau :H_1:)$ define a transfer map ${\mathcal M}_1$ and
$\exp(-\tau:H_2:)$ a transfer map ${\mathcal M}_2$, 
for a single step, the above splitting results in a second order numerical integrator
for the original Hamilton's equation as:
\begin{eqnarray}
	\zeta (\tau) & = & {\mathcal M}(\tau) \zeta(0) \nonumber \\
    & = & {\mathcal M}_1(\tau/2) {\mathcal M}_2(\tau) {\mathcal M}_1(\tau/2) \zeta(0)
	+ O(\tau^3)
	\label{map1}
\end{eqnarray}

The above numerical integrator can be extended to $4^{th}$ order 
accuracy and arbitrary even-order accuracy following
Yoshida's approach~\cite{yoshida}.
This numerical integrator Eq.~\ref{map1} will be symplectic if both the transfer map
${\mathcal M}_1$ and the transfer map ${\mathcal M}_2$ are symplectic.
A transfer map ${\mathcal M}_i$ is symplectic if and only if
the Jacobian matrix $M_i$ of the transfer
map ${\mathcal M}_i$ satisfies the following condition:
\begin{eqnarray}
M_i^T J M_i = J 
\label{symp}
\end{eqnarray}
where $J$ denotes the $6N \times 6N$ matrix given by:
\begin{equation}
	J  =   \left( \begin{array}{cc}
			0 & I \\
			-I & 0
	\end{array} \right)
\end{equation}
and $I$ is the $3N\times 3N$ identity matrix.

For the Hamiltonian in Eq.~\ref{htot}, we can choose $H_1$ as:
\begin{eqnarray}
	H_1 & = & \sum_{i=1}^{N_p} {\bf p}_i^2/2 + \sum_{i=1}^{N_p} q \psi({\bf r}_i)
\end{eqnarray}
This corresponds to the Hamiltonian of a group of charged particles 
inside an external field without mutual interaction among themselves. 
The charged particle magnetic optics method can be used to find 
the symplectic transfer map ${\mathcal M}_1$
for this Hamiltonian with the external fields from
most accelerator beam line elements~\cite{rob1,alex,mad}.

We can choose $H_2$ as:
\begin{eqnarray}
	H_2 & = & \frac{1}{2} \sum_{i=1}^{N_p} \sum_{j=1}^{N_p} q {\bar \varphi}({\bf r}_i,{\bf r}_j)
\end{eqnarray}
which includes the space-charge effect and is only a function of position. 
For the space-charge Hamiltonian $H_2({\bf r})$, the single
step transfer map ${\mathcal M}_2$ can be written as:
\begin{eqnarray}
	{\bf r}_i(\tau) & = & {\bf r}_i(0) \\
	{\bf p}_i(\tau) & = & {\bf p}_i(0) - \frac{\partial H_2({\bf r})}{\partial {\bf r}_i} \tau
	\label{map2}
\end{eqnarray}
The Jacobi matrix of the above 
 transfer map
${\mathcal M}_2$ is 
\begin{equation}
	M_2  =   \left( \begin{array}{cc}
			I & 0 \\
			L & I
	\end{array} \right)
\end{equation}
where $L$ is a $3N \times 3N$ matrix.
For $M_2$ to satisfy the symplectic condition Eq.~\ref{symp}, 
the matrix $L$ needs
to be a symmetric matrix, i.e.
\begin{equation}
	L = L^T
\end{equation}
Given the fact that $L_{ij} = \partial {\bf p}_i(\tau)/\partial {\bf r}_j =  - \frac{\partial^2 H_2({\bf r})}{\partial {\bf r}_i \partial {\bf r}_j} \tau $, the matrix $L$ will be symmetric as long as it 
is {\bf \it analytically calculated}
from the function $H_2$. 
If both the transfer map
${\mathcal M}_1$ and the transfer map ${\mathcal M}_2$ 
are symplectic, the numerical integrator Eq.~\ref{map} for multi-particle tracking will be symplectic. 

For a coasting beam, the Hamiltonian $H_2$ can be written as~\cite{rob1}:
\begin{eqnarray}
	H_2 & = & \frac{K}{2} \sum_{i=1}^{N_p} \sum_{j=1}^{N_p} \varphi({\bf r}_i,{\bf r}_j)
	\label{htot2}
\end{eqnarray}
where $K = q I/(2\pi \epsilon_0 p_0 v_0^2 \gamma_0^2)$ is the generalized perveance,
$I$ is the beam current, $\epsilon_0$ is the permittivity of vacuum,
$p_0$ is the momentum of the reference
particle, $v_0$ is the speed of the reference particle, $\gamma_0$ is
the relativistic factor of the reference particle, and $\varphi$ is the 
space charge Coulomb interaction potential.
In this Hamiltonian, the effects of the direct Coulomb 
electric potential and the
longitudinal vector potential are combined together.
The electric Coulomb potential $\varphi$ in the Hamiltonian 
$H_2$ can be obtained 
from the solution of the Poisson equation.
In the following, we assume that the coasting beam is inside a rectangular perfectly conducting pipe.
In this case, the two-dimensional Poisson's equation can be written as:
\begin{equation}
\frac{\partial^2 \phi}{\partial x^2} +
\frac{\partial^2 \phi}{\partial y^2} = - 4 \pi \rho
\label{poi2d}
\end{equation}
where
$\phi$ is the electric potential, and $\rho$ is the particle
density distribution of the beam.

The boundary conditions for the electric potential inside the rectangular 
perfectly conducting pipe are:
\begin{eqnarray}
	\label{bc1}
\phi(x=0,y) & = & 0  \\
\phi(x=a,y) & = & 0  \\
\phi(x,y=0) & = & 0  \\
\phi(x,y=b) & = & 0  
	\label{boundary}
\end{eqnarray}
where $a$ is the horizontal width of the pipe and $b$ is the vertical width
of the pipe. 

Given the boundary conditions in Eqs.~\ref{bc1}-\ref{boundary}, the electric potential $\phi$ and the
source term $\rho$ can be approximated using two sine functions as~\cite{qiang01,qiang16,gottlieb,fornberg,boyd}:
\begin{eqnarray}
	\rho(x,y)  = \sum_{l=1}^{N_l}\sum_{m=1}^{N_m} \rho^{lm} \sin(\alpha_l x) \sin(\beta_m y) \\
	\phi(x,y)  =  \sum_{l=1}^{N_l}\sum_{m=1}^{N_m} \phi^{lm} \sin(\alpha_l x) \sin(\beta_m y) 
\end{eqnarray}
where
\begin{eqnarray}
\label{rholm}
\rho^{lm}  = \frac{4}{ab}\int_0^a\int_0^b \rho(x,y) \sin(\alpha_l x) \sin(\beta_m y) \ dx dy \\
\phi^{lm}  = \frac{4}{ab}\int_0^a\int_0^b \phi(x,y) \sin(\alpha_l x) \sin(\beta_m y) \ dx dy
\end{eqnarray}
where $\alpha_l=l\pi/a$ and $\beta_m = m \pi/b$.
The above approximation
follows the numerical spectral Galerkin method since each basis functions
satisfies the boundary conditions on the wall~\cite{gottlieb,boyd,fornberg}. 
For a smooth function,
this spectral approximation has an accuracy whose numerical error
scales as $O(\exp(-cN))$ with 
$c>0$, where $N$ is the number of the basis function (i.e. mode number in each
dimension) used in the approximation.
By substituting above expansions into the
Poisson Eq.~\ref{poi2d} and making use of the orthonormal condition of the sine functions,
we obtain
\begin{eqnarray}
	\phi^{lm} & = & \frac{4 \pi \rho^{lm}}{\gamma_{lm}^2}
	\label{odelm}
\end{eqnarray}
where $\gamma_{lm}^2 = \alpha_l^2 + \beta_m^2$. 

In the simulation, the particle density distribution function $\rho(x,y)$ can be 
represented as:
\begin{eqnarray}
	\rho(x,y) & = & \frac{1}{\Delta x \Delta y N_p}\sum_{j=1}^{N_p} S(x-x_j) S(y-y_j)
\end{eqnarray}
where 
$S(x)$ is the unitless shape function (i.e. deposition function in the PIC model)
$\Delta x$ and $\Delta y$ are mesh size in each dimension. The use of the shape function
helps smooth the density function when the number of macroparticles in
the simulation is much
less than the real number of particles in the beam. 
Using the above equation and Eq.~\ref{rholm} and Eq.~\ref{odelm}, we obtain:
\begin{eqnarray}
	\phi^{lm} &  = & \frac{4 \pi}{\gamma_{lm}^2}\frac{4}{ab} \frac{1}{N_p} \sum_{j=1}^{N_p} 
	\frac{1}{\Delta x \Delta y} \int_0^a\int_0^b S(x-x_j)S(y-y_j) \sin(\alpha_l x) \sin(\beta_m y) dxdy
\end{eqnarray}
and the electric potential as:
\begin{eqnarray}
	\phi(x,y)  = {4 \pi} \frac{4}{ab} \frac{1}{N_p} \sum_{j=1}^{N_p} 
	\sum_{l=1}^{N_l} \sum_{m=1}^{N_m} 
	\frac{1}{\gamma_{lm}^2} \sin(\alpha_l x) \sin(\beta_m y) \nonumber \\
	\frac{1}{\Delta x \Delta y} \int_0^a\int_0^b S(\bar{x}-x_j)S(\bar{y}-y_j) \sin(\alpha_l \bar{x}) \sin(\beta_m \bar{y}) d\bar{x}d\bar{y}
\end{eqnarray}
The electric potential at a particle $i$ location can be obtained from the
potential as:
\begin{eqnarray}
	\phi(x_i,y_i) & = & \frac{1}{\Delta x \Delta y} \int_0^a \int_0^b \phi(x,y) S(x-x_i)S(y-y_i) dxdy  \nonumber \\
	 & = & {4 \pi} \frac{4}{ab} \frac{1}{N_p}  
	\sum_{j=1}^{N_p} 
	\sum_{l=1}^{N_l} \sum_{m=1}^{N_m} 
	\frac{1}{\gamma_{lm}^2}  \frac{1}{\Delta x \Delta y} \int_0^a \int_0^b S(x-x_i)S(y-y_i) \sin(\alpha_l x) \sin(\beta_m y)dxdy 
	\nonumber \\
	& & \frac{1}{\Delta x \Delta y} \int_0^a\int_0^b S(x-x_j)S(y-y_j) \sin(\alpha_l x) \sin(\beta_m y) dxdy 
\end{eqnarray}
where the interpolation function to the particle location is
assumed to be the same as the deposition function.

From the above electric potential, the interaction potential 
$\varphi$ between particles $i$ and $j$ can be
written as:
\begin{eqnarray}
	\varphi(x_i,y_i,x_j,y_j) & = & {4 \pi} \frac{4}{ab} \frac{1}{N_p}  
	\sum_{l=1}^{N_l} \sum_{m=1}^{N_m} 
	\frac{1}{\gamma_{lm}^2} 
	\frac{1}{\Delta x \Delta y}\int_0^a\int_0^b S(x-x_j)S(y-y_j) \sin(\alpha_l x) \sin(\beta_m y) dxdy \nonumber \\
	& &	\frac{1}{\Delta x \Delta y} \int_0^a \int_0^b S(x-x_i)S(y-y_i) \sin(\alpha_l x) \sin(\beta_m y) dxdy
\end{eqnarray}
Now, the space-charge Hamiltonian $H_2$ can be written as:
\begin{eqnarray}
	H_2 & = & 4\pi \frac{K}{2}\frac{4}{ab} \frac{1}{N_p} \sum_{i=1}^{N_p} \sum_{j=1}^{N_p} \sum_{l=1}^{N_l} \sum_{m=1}^{N_m} 
	\frac{1}{\gamma_{lm}^2}  \nonumber \\
	& &	\frac{1}{\Delta x \Delta y} \int_0^a\int_0^b S(x-x_j)S(y-y_j) \sin(\alpha_l x) \sin(\beta_m y) dxdy  \nonumber \\
	& &	       \frac{1}{\Delta x \Delta y}\int_0^a \int_0^b S(x-x_i)S(y-y_i) \sin(\alpha_l x) \sin(\beta_m y) dxdy
\end{eqnarray}

\subsection{Symplectic gridless particle model}

In the symplectic gridless particle space-charge model, 
the shape function is assumed to be a Dirac delta function and the
particle distribution function $\rho(x,y)$ can be represented as:
\begin{eqnarray}
	\rho(x,y) = \frac{1}{N_p}\sum_{j=1}^{N_p} \delta(x-x_j)\delta(y-y_j)
\end{eqnarray}

Now, the space-charge Hamiltonian $H_2$ can be written as:
\begin{eqnarray}
	H_2  = 4\pi \frac{K}{2}\frac{4}{ab} \frac{1}{N_p} \sum_{i=1}^{N_p} \sum_{j=1}^{N_p} \sum_{l=1}^{N_l} \sum_{m=1}^{N_m} 
	\frac{1}{\gamma_{lm}^2} \sin(\alpha_l x_j) 
	\sin(\beta_m y_j) \sin(\alpha_l x_i) \sin(\beta_m y_i)
\end{eqnarray}
The one-step symplectic transfer map ${\mathcal M}_2$ of the 
particle $i$ with this Hamiltonian is given as:
\begin{eqnarray}
	p_{xi}(\tau) & = & p_{xi}(0) -
	\tau 4\pi {K}\frac{4}{ab} \frac{1}{N_p} \sum_{j=1}^{N_p} 
	\sum_{l=1}^{N_l} \sum_{m=1}^{N_m} 
	\frac{\alpha_l}{\gamma_{lm}^2} 
 \sin(\alpha_l x_j) \sin(\beta_m y_j) \cos(\alpha_l x_i) \sin(\beta_m y_i)  \nonumber \\
	p_{yi}(\tau) & = & p_{yi}(0) -
	\tau 4\pi {K}\frac{4}{ab} \frac{1}{N_p} \sum_{j=1}^{N_p} 
	\sum_{l=1}^{N_l} \sum_{m=1}^{N_m} 
	\frac{\beta_m}{\gamma_{lm}^2} 
 \sin(\alpha_l x_j) \sin(\beta_m y_j) \sin(\alpha_l x_i) \cos(\beta_m y_i)
\end{eqnarray}
Here, both $p_{xi}$ and $p_{yi}$ are normalized by the reference particle momentum $p_0$.

\subsection{Symplectic particle-in-cell model}

If the deposition/interpolation shape function is not a delta function,
the one-step symplectic transfer map ${\mathcal M}_2$ of the 
particle $i$ with this space-charge Hamiltonian $H_2$ is given as:
\begin{eqnarray}
	p_{xi}(\tau) & = & p_{xi}(0) -
	\tau 4\pi {K}\frac{4}{ab} \frac{1}{N_p} \sum_{j=1}^{N_p} 
	\sum_{l=1}^{N_l} \sum_{m=1}^{N_m} 
	\frac{1}{\gamma_{lm}^2} 
	\frac{1}{\Delta x \Delta y} \int_0^a\int_0^b S(x-x_j)S(y-y_j) \sin(\alpha_l x) \sin(\beta_m y) dxdy
     \nonumber \\ 
	& &	\frac{1}{\Delta x \Delta y} \int_0^a \int_0^b 
	\frac{\partial S(x-x_i)}{\partial x_i}S(y-y_i) \sin(\alpha_l x) \sin(\beta_m y) dxdy
	\nonumber \\
	p_{yi}(\tau) & = & p_{yi}(0) -
	\tau 4\pi {K}\frac{4}{ab} \frac{1}{N_p} \sum_{j=1}^{N_p} 
	\sum_{l=1}^{N_l} \sum_{m=1}^{N_m} 
	\frac{1}{\gamma_{lm}^2} 
	\frac{1}{\Delta x \Delta y} \int_0^a\int_0^b S(x-x_j)S(y-y_j) \sin(\alpha_l x) \sin(\beta_m y) dxdy
     \nonumber \\ 
	& &	\frac{1}{\Delta x \Delta y} \int_0^a \int_0^b S(x-x_i)\frac{\partial S(y-y_i)}{\partial y_i} \sin(\alpha_l x) \sin(\beta_m y) dxdy
\end{eqnarray}
where both $p_{xi}$ and $p_{yi}$ are normalized by the reference particle momentum $p_0$,
and $S'(x)$ is the first derivative of the shape function.
Assuming that the shape function is a compact local function with respect to
the computational grid, after approximating the integral with summation on 
the grid, the above map can be rewritten as:
\begin{eqnarray}
	p_{xi}(\tau) & = & p_{xi}(0) -
	\tau 4\pi {K}\frac{4}{ab} \frac{1}{N_p} \sum_{j=1}^{N_p} 
	\sum_{l=1}^{N_l} \sum_{m=1}^{N_m} 
	\frac{1}{\gamma_{lm}^2} 
	\sum_{I'} \sum_{J'} S(x_{I'}-x_j)S(y_{J'}-y_j) \sin(\alpha_l x_{I'}) \sin(\beta_m y_{J'}) 
     \nonumber \\ 
	& &	\sum_I \sum_J \frac{\partial S(x_I-x_i)}{\partial x_i}S(y_J-y_i) \sin(\alpha_l x_I) \sin(\beta_m y_J) 
	\nonumber \\
	p_{yi}(\tau) & = & p_{yi}(0) -
	\tau 4\pi {K}\frac{4}{ab} \frac{1}{N_p} \sum_{j=1}^{N_p} 
	\sum_{l=1}^{N_l} \sum_{m=1}^{N_m} 
	\frac{1}{\gamma_{lm}^2} 
	\sum_{I'} \sum_{J'} S(x_{I'}-x_j)S(y_{J'}-y_j) \sin(\alpha_l x_{I'}) \sin(\beta_m y_{J'})
     \nonumber \\ 
	& &	\sum_I \sum_J S(x_I-x_i)\frac{\partial S(y_I-y_i)}{\partial y_i} \sin(\alpha_l x_I) \sin(\beta_m y_J) 
\end{eqnarray}
where the integers $I$, $J$, $I'$, and $J'$ denote the two dimensional computational
grid index, and the summations with respect to those indices are limited to the range
of a few local grid points depending on the specific deposition function.
If one defines the density related function $\bar{\rho}(x_{I'},y_{J'})$ on the grid as:
\begin{eqnarray}
	\bar{\rho}(x_{I'},y_{J'})  = \frac{1}{N_p} \sum_{j=1}^{N_p}  S(x_{I'}-x_j)S(y_{J'}-y_j),
	\label{2dep}
\end{eqnarray}
the above space-charge map can be rewritten as:
\begin{eqnarray}
		p_{xi}(\tau) & = & p_{xi}(0) -
	\tau 4\pi {K}\sum_I \sum_J \frac{\partial S(x_I-x_i)}{\partial x_i} S(y_J-y_i) [\frac{4}{ab} 
	\sum_{l=1}^{N_l} \sum_{m=1}^{N_m} 
	\frac{1}{\gamma_{lm}^2} 
	\nonumber \\
	& & \sum_{I'} \sum_{J'} \bar{\rho}(x_{I'},y_{J'}) \sin(\alpha_l x_{I'}) \sin(\beta_m y_{J'}) 
	 \sin(\alpha_l x_I) \sin(\beta_m y_J) ]
	\nonumber \\
	p_{yi}(\tau) & = & p_{yi}(0) -
	\tau 4\pi {K}\sum_I \sum_J S(x_I-x_i)\frac{\partial S(y_I-y_i)}{\partial y_i} [\frac{4}{ab}   
	\sum_{l=1}^{N_l} \sum_{m=1}^{N_m} 
	\frac{1}{\gamma_{lm}^2} 
	\nonumber \\
	& & \sum_{I'} \sum_{J'}	\bar{\rho}(x_{I'},y_{J'}) \sin(\alpha_l x_{I'}) \sin(\beta_m y_{J'})
	\sin(\alpha_l x_I) \sin(\beta_m y_J) ]
\end{eqnarray}
It turns out that the expression inside the bracket corresponds to the solution of potential
on grid using the charge density on grid, which can be written as:
\begin{eqnarray}
 \phi(x_I,y_J) & = & \frac{4}{ab} 
	\sum_{l=1}^{N_l} \sum_{m=1}^{N_m} 
	\frac{1}{\gamma_{lm}^2} 
	\sum_{I'} \sum_{J'} {\bar \rho}(x_{I'},y_{J'}) \sin(\alpha_l x_{I'}) \sin(\beta_m y_{J'}) 
	 \sin(\alpha_l x_I) \sin(\beta_m y_J) 
	 \label{poi2dsol}
\end{eqnarray}
Then the above space-charge map can be rewritten in a more concise form as:
\begin{eqnarray}
		p_{xi}(\tau) & = & p_{xi}(0) -
	\tau 4\pi {K}\sum_I \sum_J \frac{\partial S(x_I-x_i)}{\partial x_i} S(y_J-y_i) 
\phi(x_I,y_J)	\nonumber \\
	p_{yi}(\tau) & = & p_{yi}(0) -
	\tau 4\pi {K}\sum_I \sum_J S(x_I-x_i)\frac{\partial S(y_J-y_i)}{\partial y_i} 
	\phi(x_I,y_J)
\end{eqnarray}
In the PIC literature, a compact function such as a linear function or 
a quadratic function
is used in the simulation. For example, the derivative of 
the above TSC function can be written as:
\begin{eqnarray}
	\frac{\partial S(x_I-x_i)}{\partial x_i} & = & \left \{ \begin{array}{ll} 
		- 2(\frac{x_i - x_I}{\Delta x})/\Delta x, & |x_i - x_I| \leq \Delta x/2  \\
		(-\frac{3}{2}+\frac{(x_i-x_I)}{\Delta x})/\Delta x, & \Delta x/2 < |x_i - x_I| \leq 3/2 \Delta x, \ \  x_i > x_I \\
		(\frac{3}{2}+\frac{(x_i-x_I)}{\Delta x})/\Delta x, & \Delta x/2 < |x_i - x_I| \leq 3/2 \Delta x, \ \   x_i \leq x_I \\
		 0 & {\rm otherwise}
                    \end{array}
                  \right.
\end{eqnarray}
The same shape function and its derivative can be applied to the $y$ dimension. 

Using the symplectic transfer map ${\mathcal M}_1$ for the external field Hamiltonian 
$H_1$ from a magnetic optics code and the transfer map ${\mathcal M}_2$ for
space-charge Hamiltonian $H_2$,
one obtains a 
symplectic PIC model including the self-consistent space-charge effects.

\subsection{Benchmark of multi-particle tracking models}

\begin{figure}[!htb]
   \centering
   \includegraphics*[angle=0,width=230pt]{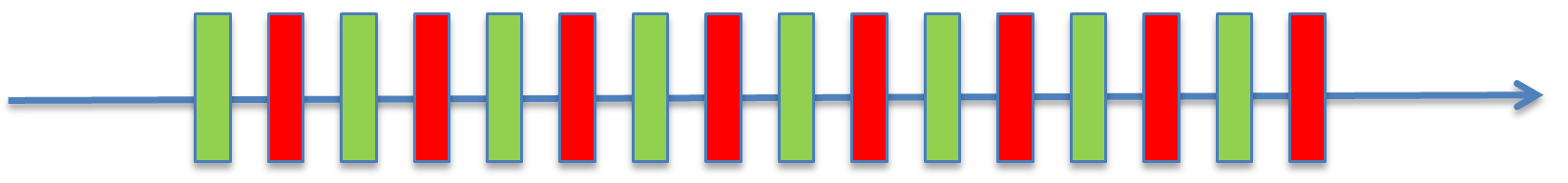}
   \caption{Schematic plot of a FODO lattice.}
   \label{fodo}
\end{figure}

The above self-consistent multi-particle tracking models were benchmarked
in an application example.
In this example, a one $GeV$ proton beam subject to
strong space-charge driven resonance transported through a  
linear periodic quadrupole focusing and defocusing (FODO) channel
inside a rectangular perfectly conducting pipe.
It was tracked including self-consistent
space-charge effects for several hundred thousands of
lattice periods using the symplectic gridless model, the symplectic
particle-in-cell model and the nonsymplectic particle-in-cell model. 
A schematic plot of the lattice is shown in Fig.~\ref{fodo}.
It consists of a $0.1$ m focusing quadrupole magnet and a $0.1$ m defocusing 
quadrupole magnet within a single period. 
The total length of the period is $1$ meter.
The zero current phase advance through one lattice period is $85$ degrees.
The current of the beam is $450$ A and the depressed phase advance
is $42$ degrees. Such a high current drives the beam across
the $4^{th}$ order collective instability~\cite{ingo83}.
The initial transverse normalized emittance of the proton beam is $1$ $\mu m$ 
with a 4D Gaussian distribution. 

\begin{figure}[!htb]
   \centering
   \includegraphics*[angle=0,width=230pt]{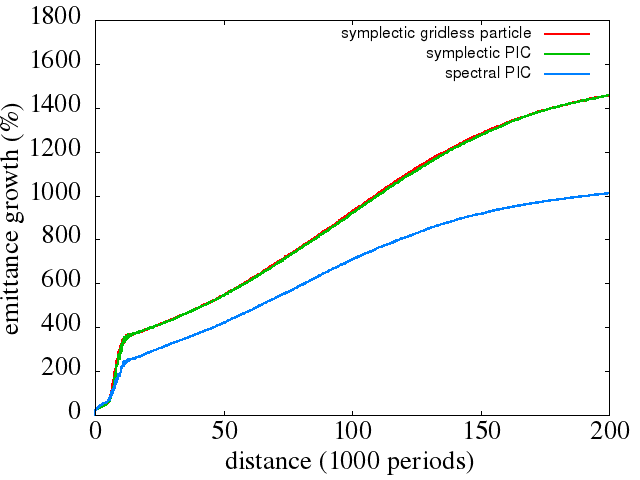}
   \caption{
	   Four dimensional emittance growth evolution from the
	symplectic gridless particle model (red), the symplectic PIC model (green),
	and the nonsymplectic spectral PIC (blue).}
   \label{emtfodo}
\end{figure}

Fig.~\ref{emtfodo} shows the four dimensional emittance growth
$(\frac{\epsilon_x}{\epsilon_{x0}}\frac{\epsilon_y}{\epsilon_{y0}}-1)\%$
evolution through $200,000$ lattice periods from the symplectic gridless
particle model, from the symplectic PIC model, and from the nonsymplectic
spectral PIC model. These simulations used about $50,000$ macroparticles
and $15\times 15$ modes in the spectral Poisson solver.
In both PIC models, $257\times 257$ grid points are used to obtain the density
distribution function on the grid. 
The perfectly conducting pipe has a square shape with an aperture size
of $10$ mm.
It is seen that the symplectic PIC model and the symplectic gridless particle
model agrees with each other very well. The nonsymplectic spectral PIC model 
yields significantly
smaller emittance growth than those from the two symplectic methods,
which might result from the numerical damping effects in the nonsymplectic
integrator.
The fast emittance growth within the first $20,000$ periods is 
caused by the space-charge driven $4^{th}$ order collective instability.
The slow emittance growth after $20,000$ periods might be due to numerical
collisional effects.

\begin{figure}[!htb]
   \centering
   \includegraphics*[angle=0,width=230pt]{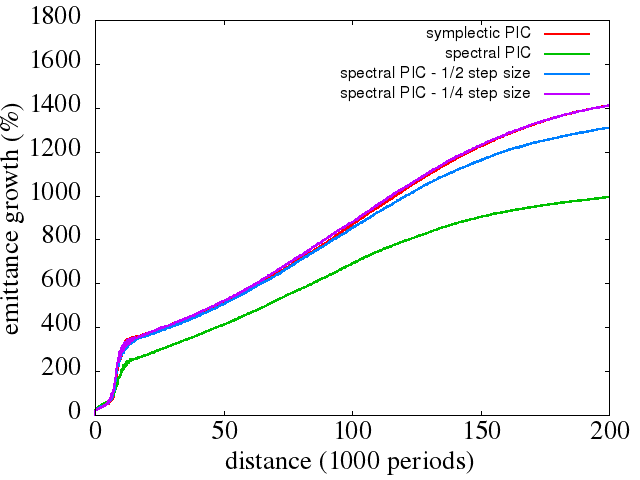}
   \caption{
	   Four dimensional emittance growth evolution from the
	symplectic PIC model (red), the nonsymplectic
	spectral PIC with the nominal step size (green), the spectral PIC
	with half of the nominal step size (blue), and the spectral PIC
	with one quarter of the nominal step size (pink).}
   \label{emtfodo2}
\end{figure}

The accuracy of the nonsymplectic PIC model can be improved with finer
step size. Fig.~\ref{emtfodo2} shows the 4D emittance growth
evolution from the symplectic PIC model and those from the nonsymplectic PIC
model with the same nominal step size, from the nonsymplectic PIC model
with one-half of the nominal step size, and from the nonsymplectic PIC
model with one-quarter of the nominal step size. It is seen that as
the step size decreases, the emittance growth from the nonsymplectic PIC
model converges towards that from the symplectic PIC model.

The computational complexity of the PIC model is proportional to
the $O(N_{grid}log(N_{grid})+N_p)$, where $N_{grid}$ and $N_p$ are total number
of computational grid points and macroparticles used in the simulation.
The computational complexity of the symplectic gridless particle model is
proportional to the $O(N_{mode}N_p)$, where $N_{mode}$ is the total number
of modes for the space-charge solver.
On a single processor computer, with a large number of macroparticles used,
the symplectic PIC model is computationally
more efficient than the symplectic gridless particle model.
However, on a massive parallel computer, the scalability of the
symplectic PIC model may be limited by the challenges of 
work load balance among multiple processors and communication associated
with the grided based space-charge solver and the particle manager~\cite{qiangcpc}. 
The symplectic gridless particle model has regular data structure
for perfect work load balance and small
amount of communication associated with the space-charge solver.
It scales well on both multiple processor Central Processing Unit (CPU)
computers and multiple Graphic Processing Unit (GPU) computers~\cite{liu}.

\section*{Acknowledgements}
This work was supported by the U.S. Department of Energy under Contract No. DE-AC02-05CH11231 and
used computer resources at the National Energy Research
Scientific Computing Center.

\end{document}